\documentclass[floatfix, noshowpacs, preprintnumbers, twocolumn, amsmath, amssymb, aps, prl, superscriptaddress]{revtex4}

\usepackage{graphicx, xcolor}
\usepackage[caption=false]{subfig}
\usepackage{soul}

\newcommand{\bra}[1]{\left< #1 \right|} 
\newcommand{\ket}[1]{\left| #1 \right>}

\begin{document}

\title{Observation of Majorization Principle for quantum algorithms\\ via 3-D integrated photonic circuits}

\author{Fulvio Flamini}
\affiliation{Dipartimento di Fisica, Sapienza Universit\`{a} di Roma,
Piazzale Aldo Moro 5, I-00185 Roma, Italy}

\author{Niko Viggianiello}
\affiliation{Dipartimento di Fisica, Sapienza Universit\`{a} di Roma,
Piazzale Aldo Moro 5, I-00185 Roma, Italy}

\author{Taira Giordani}
\affiliation{Dipartimento di Fisica, Sapienza Universit\`{a} di Roma,
Piazzale Aldo Moro 5, I-00185 Roma, Italy}

\author{Marco Bentivegna}
\affiliation{Dipartimento di Fisica, Sapienza Universit\`{a} di Roma,
Piazzale Aldo Moro 5, I-00185 Roma, Italy}

\author{Nicol\`o Spagnolo}
\affiliation{Dipartimento di Fisica, Sapienza Universit\`{a} di Roma,
Piazzale Aldo Moro 5, I-00185 Roma, Italy}


\author{Andrea Crespi}
\affiliation{Istituto di Fotonica e Nanotecnologie, Consiglio Nazionale delle Ricerche (IFN-CNR), 
Piazza Leonardo da Vinci, 32, I-20133 Milano, Italy}
\affiliation{Dipartimento di Fisica, Politecnico di Milano, Piazza Leonardo da Vinci, 32, I-20133 Milano, Italy}

\author{Giacomo Corrielli}
\affiliation{Istituto di Fotonica e Nanotecnologie, Consiglio Nazionale delle Ricerche (IFN-CNR), 
Piazza Leonardo da Vinci, 32, I-20133 Milano, Italy}

\author{Roberto Osellame}
\affiliation{Istituto di Fotonica e Nanotecnologie, Consiglio Nazionale delle Ricerche (IFN-CNR), 
Piazza Leonardo da Vinci, 32, I-20133 Milano, Italy}
\affiliation{Dipartimento di Fisica, Politecnico di Milano, Piazza Leonardo da Vinci, 32, I-20133 Milano, Italy}

\author{Miguel Angel Martin-Delgado}
\affiliation{Departamento de F\'{\i}sica Te\'orica I, Universidad Complutense, 28040 Madrid, Spain}

\author{Fabio Sciarrino}
\email{fabio.sciarrino@uniroma1.it}
\affiliation{Dipartimento di Fisica, Sapienza Universit\`{a} di Roma,
Piazzale Aldo Moro 5, I-00185 Roma, Italy}

\begin{abstract}
The Majorization Principle is a fundamental statement governing the dynamics of information processing in optimal and efficient quantum algorithms.
While quantum computation can be modeled to be reversible, due to the unitary evolution undergone by the system, these quantum algorithms are conjectured to obey a quantum arrow of time dictated by the Majorization Principle:
the probability distribution associated to the outcomes gets ordered step-by-step until achieving the result of the computation. Here we report on the experimental observation of the effects of the Majorization Principle for two quantum algorithms, namely the quantum fast Fourier transform and a recently introduced validation protocol for the certification of genuine many-boson interference. The demonstration has been performed by employing integrated 3-D photonic circuits fabricated via femtosecond laser writing technique, which allows to monitor unambiguously the effects of majorization along the execution of the algorithms. The measured observables provide a strong indication that the Majorization Principle holds true for this wide class of quantum algorithms, thus paving the way for a general tool to design new optimal algorithms with a quantum speedup.
\end{abstract}
\maketitle

\textit{Introduction ---} Quantum computation holds the promise to greatly improve the capabilities of computational platforms relying on the laws of classical physics \cite{nielsen_chuang}. Such potentiality arises from the combination of both an exponential storage capability and a dynamical parallel processing of the unitary time evolutions. However, the unprecedented massive computational resource offered by the parallel processing alone is doomed to failure, due to the non-deterministic nature of any measurement process. Thus, quantum algorithms have been properly tailored to exploit the power hidden in quantum resources challenging this limitation \cite{CEMM_98}.
While quantum correlations are considered to be the fundamental physical resource responsible for the higher efficiency of quantum algorithms, we still ignore how to manage them to effectively produce new quantum protocols \cite{Vedral2010}.
This situation is in sharp contrast with the theory of classical algorithms, where there exist well-known strategies to devise new algorithms starting from those already available \cite{delgado_02}. However, while a complete picture of the principles governing the design of new quantum algorithms is still lacking, we can yet control the evolution to guarantee that a new alleged quantum algorithm is truly efficient. Such criterion may arguably be provided by the Majorization Principle (MP) \cite{MP1,MP2,MP3}. So far, indeed, all time evolutions ultimately belong to two categories. Classical evolutions can be described by the Principle of Least Action \cite{goldstein}: trajectories must obey local constraints so that the Action remains stationary at each point of the geodesic. On the contrary, a global description for quantum evolutions requires to sum over all classical trajectories weighted by the exponential of the phase-Action \cite{feynman_hibbs}.
MP-constrained evolutions represent a synthesis of these two typologies, lying between classical observables and purely quantum processing. Specifically, the MP is believed to provide a necessary condition that must hold to produce an optimal algorithm with a quantum speedup.

\vspace{1em}
Here we report on the experimental observation of the Majorization Principle, acting along the execution of both a Quantum Fourier Transform (QFT) routine and a recently introduced quantum validation protocol aimed at certifying genuine many-boson interference \cite{Tichy2013}. The importance of these two operations in the context of quantum algorithms make them a perfect test-bed for the experimental demonstration of the occurrence of the MP. The observation has been carried out by implementing the protocols on 3D integrated photonic circuits realized via femtosecond laser writing technique on alumino-borosilicate substrates \cite{Osellame2003,gattass2008flm,Spagnolo13,Crespi2016}. This fabrication procedure presents the unique advantage of permitting interferometric architectures with 3-dimensional topology. The latter feature enabled the capability of decomposing the action of the two protocols into discrete steps, through which it has been possible to observe the MP.

\begin{figure*}[ht!]
\includegraphics[width=17.5cm, height=5.2cm]{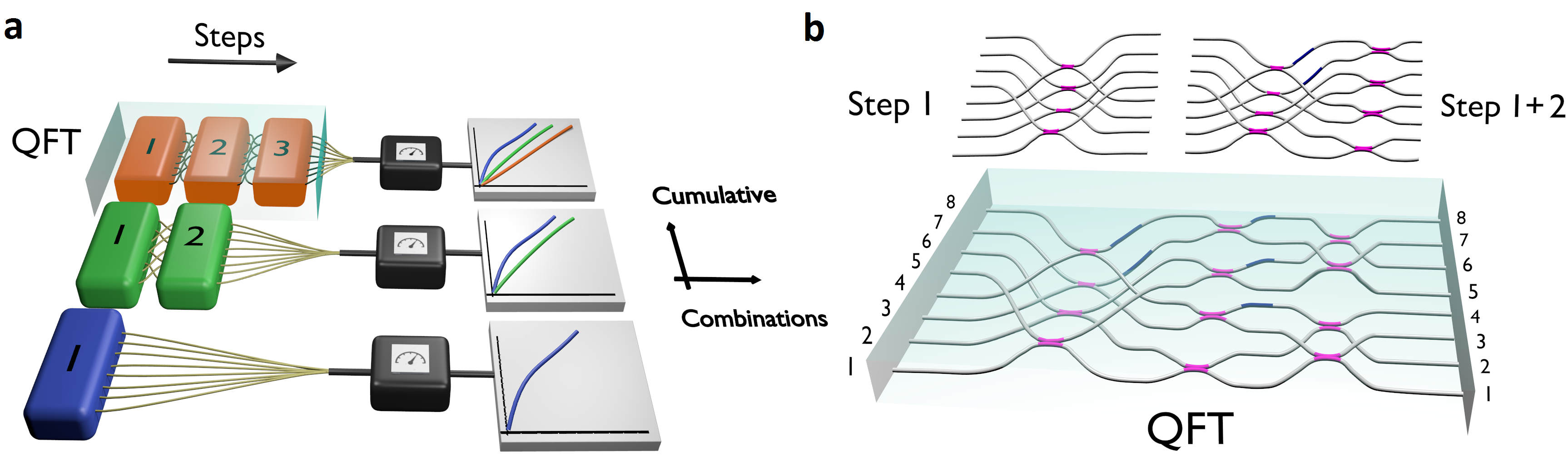}
\caption{\textbf{Majorization Principle in quantum Fourier transform.} (\textbf{a}) Conceptual scheme for the experimental observation of the Majorization Principle in (\textbf{b}) an 8-dimensional quantum Fourier transform: by implementing the QFT with its fast architecture, it is possible to decompose the evolution in three steps, corresponding to the three layers of beamsplitters (purple) and phase shifters (blue) between which a step-by-step majorization is observed.}
\label{fig:Concept}
\end{figure*}

The Majorization Principle arises with respect to the probability distributions of all possible outcomes of an algorithm, updated while advancing throughout each step. Given two probability distributions ($\vec{x}$,  $\vec{y}$), let ($\vec{x}^{\downarrow}$, $\vec{y}^{\downarrow}$) be the same vectors with their components sorted in decreasing order. We say that $\vec{y}^{\downarrow}$ majorizes $\vec{x}^{\downarrow}$ ($\vec{y}^{\downarrow} \succ \vec{x}^{\downarrow}$) if and only if
 \begin{equation}
\sum_{i=1}^k x_i^{\downarrow} \le \sum_{i=1}^k y_i^{\downarrow} ,    \qquad   \forall \:  k \in \{1, ..., d\}
\label{MP}
\end{equation}


The concept of majorization can be extended in a natural way to quantum algorithms. Let $| \psi^{(s)} \rangle$  represent the state of the register of a quantum computer at a step \textit{s} of the algorithm. We can associate to $| \psi^{(s)} \rangle$ a vector of  probabilities $p^{(s)}$ by writing the register in the computational basis $|  j \rangle $, in such a way that  $p^{(s)} _{j} = |\langle j  |  \psi^{(s)} \rangle  |^{2} $.
Consequently, a quantum algorithm is said to undergo a direct (reversed) majorization if and only if ${p^{(s)}}^{\downarrow} \prec {p^{(s+1)}}^{\downarrow}$ 
(${p^{(s)}}^{\downarrow} \succ 
{p^{(s+1)}}^{\downarrow} $) for all steps $s\;$ \cite{MP2}. An intuitive reason for the physical connection between quantum processing and direct (reversed) majorization is that of a neat flux of probability towards (away from) the result of the computation, making the probability distribution steeper (flatter) throughout the whole algorithm.\\The principle can now be stated as follows \cite{MP1}:

\vspace{1em}
\textbf{Majorization Principle:} \textit{In all optimal and efficient quantum algorithms, the set of sorted probabilities associated to the quantum register must obey either a direct or a reverse step-by-step majorization.}

\vspace{1em}

\noindent All known quantum algorithms which are both optimal and efficient, i.e. with a quantum speedup over the best classical algorithm, have been proven to satisfy the conjectured MP with a direct or reverse majorization \cite{MP3}. Remarkably, similar majorization constraints have already found applications in highlighting arrows in several other physical processes \cite{marshall_olkin,vidal_cirac_02}. Similarly, the MP promises to represent the arrow which operates within optimal and efficient quantum algorithms.

\noindent The validity of the principle has been proved theoretically for both Grover-like \cite{grover_96} and phase estimation-like \cite{shor_97} algorithms. Further optimal algorithms studied include a variant of the Berstein-Vazirani algorithm \cite{BV97}, a set of quantum adiabatic algorithms \cite{Farhi2000} and a quantum random walk algorithm \cite{Childs2002}. For all such instances, quantum speedups over the classical state of the art were always found to be associated to a step-by-step majorization, while non-efficient computations did not. The case of the Berstein-Vazirani algorithm is of even greater interest, since no entanglement is created along the computation, while majorization is indeed verified \cite{MP3}.  Thus, a strong evidence exists that the MP will represent a fundamental tool for the design of future efficient quantum algorithms. The goal of this paper is to provide experimental evidence that this statement holds true for two quantum algorithms, the quantum Fourier transform and a recently proposed validation protocol.

\vspace{1em}
\textit{Majorization Principle in a fast QFT ---} The class of phase-estimation algorithms, which includes Shor's factorization and discrete logarithms \cite{shor_97}, is of particular importance for the exponential speedup over the best available classical equivalents. Such quantum speedup is ultimately rooted in the efficient processing of the QFT routine, whose \textit{m}-dimensional unitary evolution $U_{m}^{\operatorname{QFT}}$ 
is given by 
$ | l \rangle \rightarrow  \frac{1}{\sqrt{m}} \sum_{q=0}^{m-1} \, e^{2 \pi i \, \frac{ l q}{m}} \, | q \rangle$.

In this article, we report on the experimental observation of the MP in the case of the QFT, where the routine is encoded in the optical modes. The corresponding transformation has been implemented on a photonic platform by adopting an efficient scheme developed by Barak and Ben-Aryeh \cite{Barak2007} (BB) to minimize the number of optical elements required. This scheme represents the quantum analogue of the Fast Fourier Transform (qFFT), the well-known classical algorithm to efficiently calculate the discrete Fourier transform. By adopting this approach, valid for transformations of dimension $m=2^p$, the necessary number of beamsplitters and phase shifters is significantly reduced to $(m/2) \log m$ \cite{Barak2007}, from the $O(m^2)$ elements needed for the most general decompositions \cite{Clements16}. The qFFT has been realized on photonic integrated interferometers taking advantage of the 3-D capabilities of femtosecond laser writing \cite{Osellame2003, gattass2008flm}, which allows to arrange the waveguides in arbitrary and fully-scalable three-dimensional structures \cite{Spagnolo13,Crespi2016}. More in particular, the step-by-step reversed majorization can be directly monitored thanks to the sequential structure that naturally emerges from the BB decomposition, as shown in Fig.\ref{fig:Concept}. The observation has been carried out by injecting single-photon Fock states into three 8-mode integrated interferometers  $\{I_1,I_2,I_3\}$ which corresponds to partial implementations of the qFFT protocol, with different degrees of completion. The number of fabricated interferometers $I_s$, each consisting of $s$ layers of beamsplitters and phase shifters, corresponds to the number of layers in the decomposition of an 8-dimensional QFT. The last interferometer $I_3$ performs the complete 8-mode qFFT, where one photon encodes 3 qubits over the optical modes. The effective unitaries implemented by the physical interferometers, which differ from the ideal ones due to unavoidable experimental imperfections, have then been reconstructed. The reconstruction process has been performed by exploiting a-priori knowledge on the architecture, to estimate the transmissivities of the directional couplers and the relative phases in the phase shifters \cite{Crespi2015}. Parameters have been retrieved by minimizing a suitable $\chi^2$ function with the single-photon and two-photon measurements. The fidelities $\mathcal{F}_s$ between the reconstructed transformations in the $I_s$ and the ideal unitaries obtained with the decomposition are  $\mathcal{F}_1=0.9954 \pm 0.0002$, $\mathcal{F}_2=0.9921 \pm 0.0005$ and $\mathcal{F}_3=0.9527 \pm 0.0006$, thus confirming the quality of the fabrication process.  The errors have been estimated with a Monte Carlo simulation, by sampling 1000 sets of new experimental data normally distributed around the ones measured.

\begin{figure}[ht!]
\includegraphics[width=0.4\textwidth]{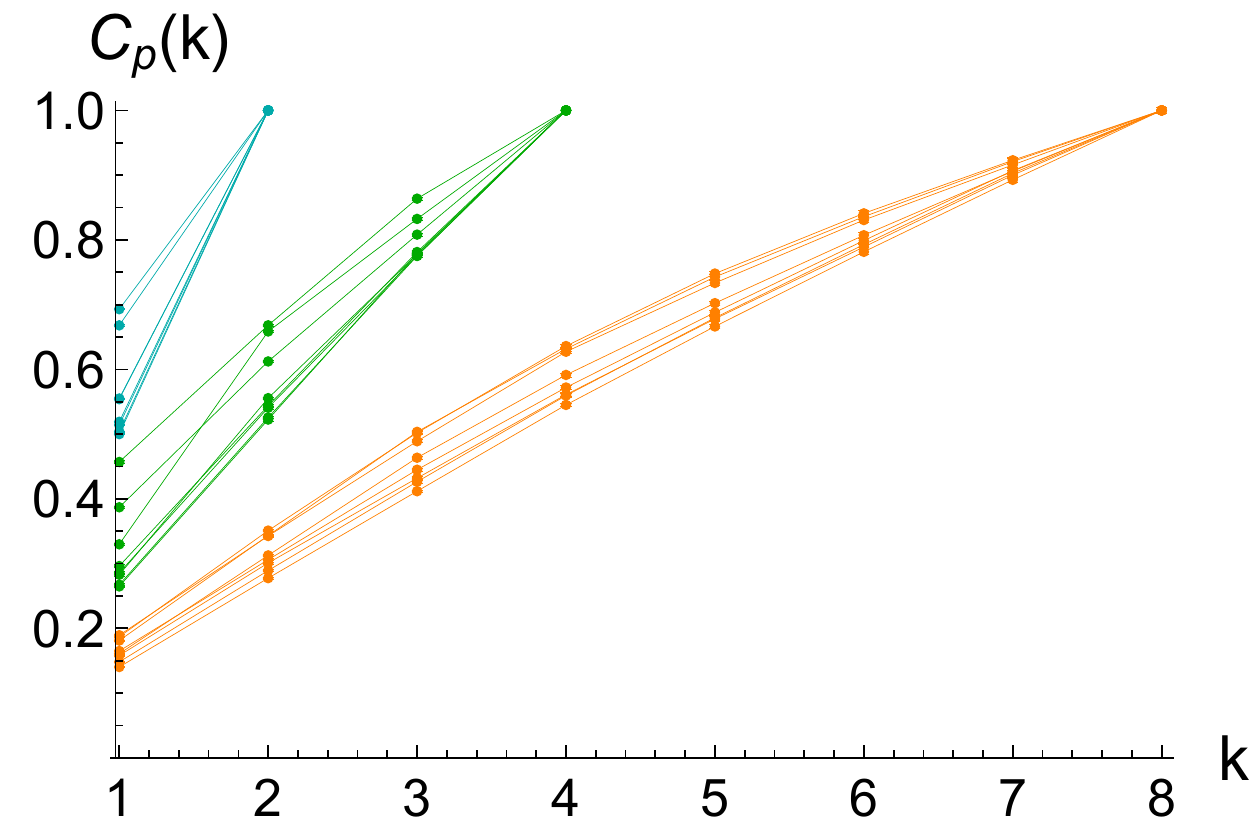}
\caption{Lorenz diagrams for a QFT Majorization experiment, with a single photon encoding three qubit on the first layer $I_1$ (blue), first and second layer $I_2$ (green) and complete structure $I_3$ (orange) of the 8-dimensional Fourier interferometer.  For each intermediate $I_s$, eight diagrams are shown relative to all possible single-photon input states. Each curve represents the partial cumulative probabilities $C_p(k)$ for the $k$ most probable outcomes. Error bars are estimated with a Monte Carlo simulation, to take into account the sorting procedure. }
\label{ResultsQFT}
\end{figure}

For each input state $i$ and each partial transformation $I_s$, the output probability distributions $p_i$ have been retrieved for the eight output states. The most convenient tool to convey the validity of the Majorization Principle is then offered by the Lorenz curve, a continuous piecewise linear function representing the partial cumulative $C_p(k) = \sum_{i=1}^k p_i^{\downarrow}$ for the $k$ most probable outcomes. For the MP to be satisfied, the curves $C_p$ at each step of the QFT must not cross, due to the inequality \eqref{MP}. As shown in Fig.\ref{ResultsQFT}, a step-by-step reversed majorization is then observed between the output probability distributions of the three interferometers $I_s$, i.e. by comparing $C_{p^{(1)}}(k)$, $C_{p^{(2)}}(k)$ and $C_{p^{(3)}}(k)$ according to (1).

\begin{figure}[ht!]
\includegraphics[width=0.46\textwidth]{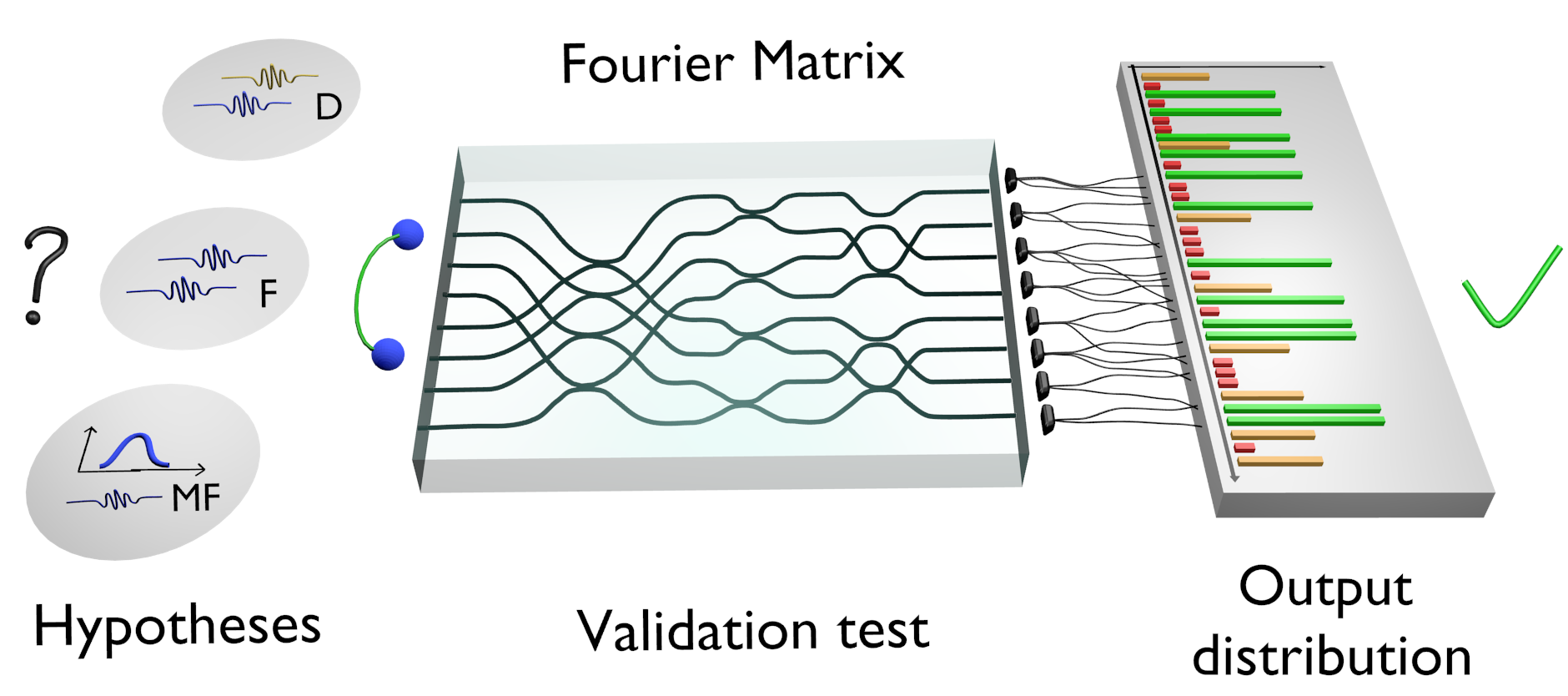}
\caption{Scheme of the validation protocol. The algorithm certifies full many-boson interference of Fock states (F) in a Fourier interferometer, against the alternative hypotheses of Distinguishable (D) and Mean-Field (MF) states. }
\label{Validation_scheme}
\end{figure}

\begin{figure*}[ht!]
	\includegraphics[width=0.7\textwidth]{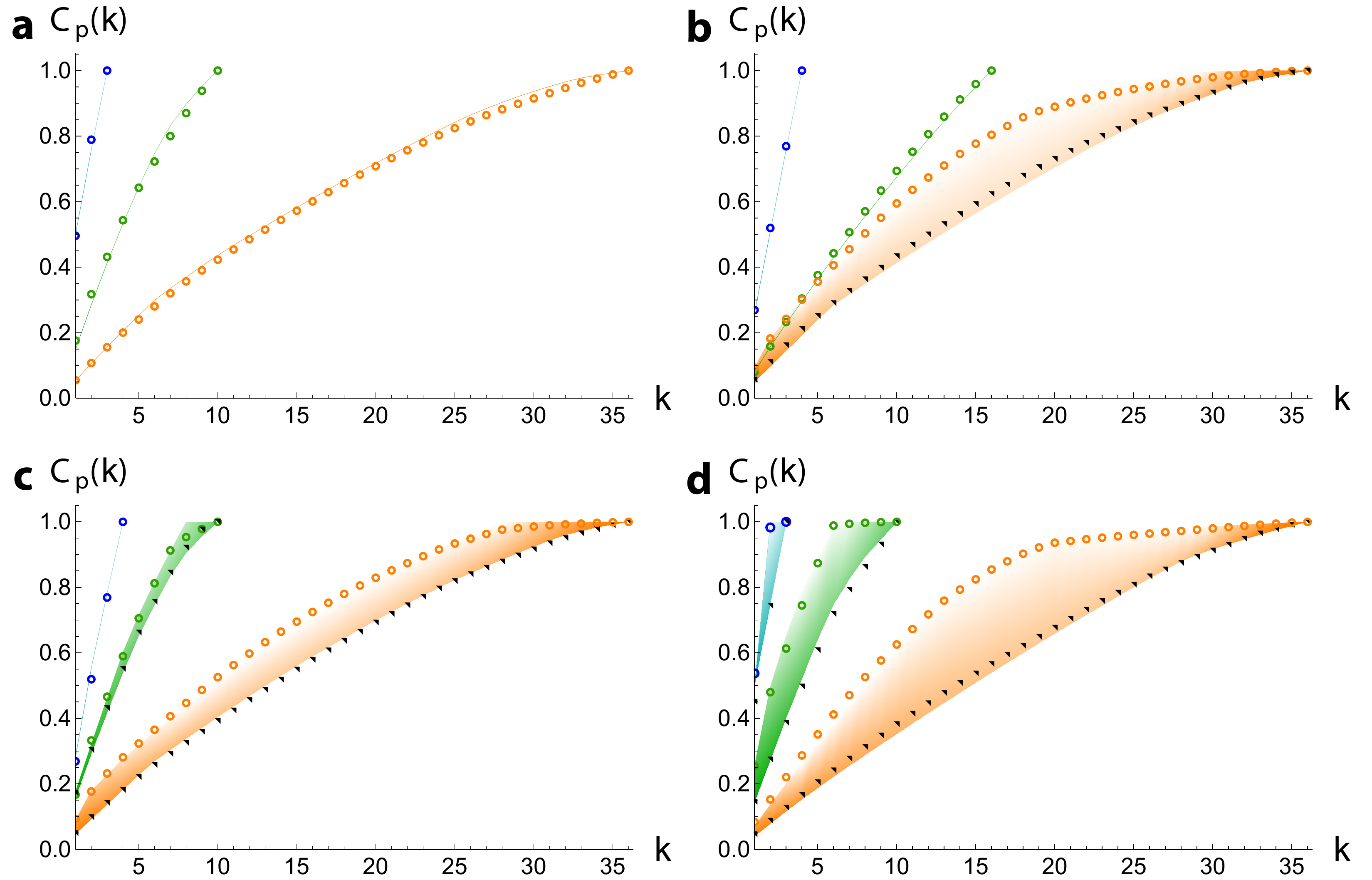}
	\caption{
	 Lorenz diagrams for a two-photon Majorization experiment on the first layer $I_1$ (blue), first and second layer $I_2$ (green) and complete structure $I_3$ (orange) of the 8-dimensional Fourier interferometer for the validation algorithm. \textbf{a}) Two photons in the same input mode (5,5). The distribution is the product of two QFT acting on the single photons. \textbf{b, c}) Input modes whose sum is odd (5,6) or even (5,7) respectively. \textbf{d}) Cyclic input (2,6) for the validation algorithm. Each curve is obtained by calculating the partial cumulative probabilities $C_p(k)$ for the $k$ most probable outcomes in the case of distinguishable (black triangles) and indistinguishable (circles) photons. Shaded areas are included within the curves corresponding to fully indistinguishable photons (lighter regions) and to fully distinguishable photons (darker regions), as expected from the reconstructed unitary transformations. Error bars, smaller than the markers, have been estimated via a Monte Carlo simulation to take into account the sorting procedure.}
\label{ResultsValidation}
\end{figure*}

\vspace{1em}
\textit{Majorization Principle in a validation protocol ---}  Quantum computation aims at developing algorithms able to outperform the classical counterparts on specific tasks. However, in this sought-after regime of a quantum supremacy, where standard computers no longer can check the results of a quantum device, the need for a quantum validation protocol becomes urgent and fundamental. This necessity arises prominently in the context of Boson Sampling experiments
\cite{Broome2013,Spring2013,Till2012,Crespi2012,Spagnolo2013a,Carolan2013a,Bentivegna2015,Carolan2015}, 
specialized devices whose task is to provide a first evidence of this future quantum supremacy \cite{AA10}. In this direction, various protocols have been developed \cite{AA13, Tichy2013, Crespi2015, Gogolin2015, Friburgo2016, Spagnolo2016} and implemented \cite{Spagnolo2013a,Carolan2013a, Carolan2015,Crespi2016} to certify their correct functioning against undesired alternatives \cite{Gogolin2013}. One of these protocols, recently developed by Tichy et al. \cite{Tichy2013}, allows to efficiently certify the source of a Boson Sampling experiment,  ruling out alternative hypotheses which would yield output probability distributions similar to that with fully  indistinguishable photons. In particular it was observed that, for symmetric input states of specific interferometers, quantum many-particle interference may determine the suppression of a large number of output combinations in a way efficiently predictable \cite{Tichy2013}, i.e. without having to go through the calculation of a permanent, which is at the core of the computational complexity of the Boson Sampling problem \cite{AA10}. Specifically, denoting with $\ket{R}=\ket{{r}_1{r}_2...{r}_m}= \hat{a}_{1}^{\dag {r}_1} \hat{a}_{2}^{\dag {r}_2}...\hat{a}_{m}^{\dag {r}_m} \ket{0}$ a generic input state with ${r}_i$ indistinguishable photons in the mode $i$ of the $m$-mode interferometer, the probability $p^{(U)} _{S,T}$ of having a certain output configuration $\ket{T}$, given one in the input $\ket{S}$, is given by
\begin{equation}
p^{(U)} _{S,T}=\big{|} \bra{T} U \ket{S}\big{|}^2=\frac{| \textup{Per}(U_{ST}) |^2}{  \prod_{i=1}^m s_i ! \; t_i !}
\label{permanent}
\end{equation}
\noindent being $U_{ST}$ the submatrix obtained by repeating $s_i$ ($t_i$) times the column $i$ (row $j$) of $U$, and being $\textup{Per}(M)$ the permanent of the matrix M $\,$ \cite{Scheel04}.
\noindent Indeed, let us consider a $n^p$-dimensional Fourier interferometer described by $(U_{n^{p}}^{\operatorname{F}})_{l,q}=(n^{p})^{-1/2}\, e^{2 \pi i \, \frac{ l q}{n^{p}}}$ . When injected with cyclic input states, i.e. $n$-photon Fock states distributed over the input modes satisfying $j^{(a,b)}=b+(a-1) \, n^{p-1}$, with  $a\in\{1, \ldots, n\}$, $b\in\{1, \ldots, n^{p-1}\}$ and $p \in \mathbb{N}$, they all result in the suppression of the output combinations which do not satisfy the relation $ \mod \!\!\left(\sum_{l=1}^n m_l, \,n\right ) = 0 $, being $m_l$ the output mode of the $l^{th}$ photon \cite{Tichy2013,Carolan2015,Crespi2016}.

The efficiency and scalability of this algorithm are crucial features for its application in a hard to simulate regime. Hence, we expect the MP to be always satisfied while certifying genuine many-photon interference for the cyclic input states.
The experiment was carried out by injecting two-photon states into the three 8-mode integrated interferometers  $\{I_1,I_2,I_3\}$ implementing partial instances of the qFFT. According to the validation test, a suppression of specific output configurations was expected due to the interference of symmetric states. This effect was indeed observed by measuring all $\binom{8+2-1}{2}=36$ two-photon coincidence events at the output of each $I_s$ for a given cyclic state, to retrieve the scattering probabilities $ p^{(s)} _{i,j} $ of having two photons in the output modes $(i,j)$. For the MP to be observed, the whole set of outcomes has to be recorded: this requirement involved the measurement of the eight bunching events $(i,i)$, i.e. when two-photon exited from the same output mode. This measurement was carried out by adding, at the end of the fiber array coupled to the output of the interferometer, additional fiber optic splitters to redirect the bunched photons in two separate detectors. A detection system was then able to register all the one-to-one coincidences between any number of firing detectors.
For all three $I_s$, the probability distributions $ p^{(s)}$ of 4 two-photon input states have been measured and plugged into \eqref{MP} to test the validity of the MP. All 36 patterns for the partial cumulative probabilities $C_p(k)$ can in fact be divided in four distinct classes \cite{SI}, as shown in Fig.\ref{ResultsValidation}. Non-crossing curves are expected for Fig.\ref{ResultsValidation}a, since the probability distribution is the product of two single-photon QFT. Furthermore, we observe in Fig.\ref{ResultsValidation}b-c that non-crossing curves are present also for non-cyclic input states, which are not employed in the validation protocol \cite{Tichy2013}. This latter observation confirms that the occurrence of the MP does not imply optimality, since the principle does not provide a sufficient condition. Finally, the non-crossing Lorenz curves relative to the cyclic input state of the validation protocol (Fig.\ref{ResultsValidation}d), manifesting that  $p^{(s)} \succ  p^{(s+1)}$ at each stage of the evolution, confirm the operation of the principle along the quantum algorithm.

\vspace{1em}
\textit{Discussion ---} We have reported on the experimental demonstration of the Majorization Principle for two efficient quantum algorithms, the quantum Fourier transform and a recently proposed protocol for validating true many-boson interference. The observation was carried on an integrated photonic platform, realized by adopting a novel 3-D architecture fabricated via femtosecond laser writing technique. Single photon and two-photon measurements on an 8-dimensional Fourier interferometer have shown the occurrence of the  Majorization Principle all along the two quantum protocols, by exploiting a fast decomposition of the evolution in discrete steps. 
The results obtained provide experimental evidence for the Majorization Principle, making it a promising guide for devising new quantum algorithms with a speedup over their corresponding classical counterpart. The good agreement with the expected distributions highlight the quality of the 3-D capabilities of femtosecond laser-writing, thus confirming it as an effective tool for addressing broader investigations on photonic platforms.

\vspace{1em}

\textbf{Acknowledgements.} This work was supported by the ERC-Starting Grant 3DQUEST (3D-Quantum Integrated Optical Simulation; grant agreement no. 307783): http://www.3dquest.eu, and by the Spanish MINECO grant FIS2015-67411, FIS2012-33152, the CAM research consortium QUITEMAD+ S2013/ICE-2801, and U.S. Army Research Office through grant W911NF-14-1-0103 for partial financial support.

\bigskip

\noindent web site: http://www.quantumlab.it/


\begin{thebibliography}{10}


\bibitem{nielsen_chuang}
\bibinfo{author}{ M. A. Nielsen and I. L. Chuang,}
\newblock \bibinfo{title}{\textit{Quantum Computation and Quantum Information}.}
(\bibinfo{year}{Cambridge University Press, New York, 2010}).


\bibitem{CEMM_98}
\bibinfo{author}{R. Cleve, A. Ekert, C. Macchiavello, and M. Mosca,}
\newblock \bibinfo{title}{Quantum algorithms revisited.}
(Proc. R. Soc. A, 1998)


\bibitem{Vedral2010}
\bibinfo{author}{V. Vedral,}
\newblock \bibinfo{journal}{Found. Phys.} \textbf{\bibinfo{volume}{82}}, 8 (\bibinfo{year}{2010}).



\bibitem{Lloyd99}
\bibinfo{author}{S. Lloyd,}
\newblock \bibinfo{journal} {Phys. Rev. A }\textbf{\bibinfo{volume}{61}} 
(\bibinfo{year}{1999}).

\bibitem{delgado_02}
\bibinfo{author}{A. Galindo and M. A. Martin-Delgado,}
\newblock \bibinfo{journal}{Rev. Mod. Phys.}
\textbf{\bibinfo{volume}{74}}, 347
(\bibinfo{year}{2002}).



\bibitem{MP1}
\bibinfo{author}{J. I. Latorre and M. A. Martin-Delgado,}
\newblock \bibinfo{journal}{Phys. Rev. A}
\textbf{\bibinfo{volume}{66}}, 022305
(\bibinfo{year}{2002}).


\bibitem{MP2}
\bibinfo{author}{R. Orus,  J. I. Latorre, and M. A. Martin-Delgado,}
\newblock \bibinfo{journal}{Quant. Inf. Proc.}
\textbf{\bibinfo{volume}{1}} (4), 283
(\bibinfo{year}{2002}).


\bibitem{MP3}
\bibinfo{author}{R. Orus,  J. I. Latorre, and M. A. Martin-Delgado,}
\newblock \bibinfo{journal}{EPJ D}
\textbf{\bibinfo{volume}{29}}, 119
 (\bibinfo{year}{2004}).

\bibitem{goldstein}
\bibinfo{author}{H. Goldstein, C. P. Poole, and J. L. Safko,}
\newblock \bibinfo{title}{\textit{Classical Mechanics (3rd Edition)}}
(\bibinfo{year}{Addison-Wesley, 2001}).


\bibitem{feynman_hibbs}
\bibinfo{author}{ R. P. Feynman and A. R. Hibbs,}
\newblock \bibinfo{title}{\textit{Quantum Mechanics and Path Integrals}}
\newblock (\bibinfo{year}{McGraw Hill: 1965 Dover Publications}).




\bibitem{Tichy2013}
\bibinfo{author}{M. C. Tichy, K. Mayer, A. Buchleitner, and K. M\o lmer},
\newblock \bibinfo{journal}{Phys. Rev. Lett.}
\textbf{\bibinfo{volume}{113}}, 020502
(\bibinfo{year}{2014}).
  
 


\bibitem{Osellame2003}
\bibinfo{author}{R. Osellame, S. Taccheo, M. Marangoni, R. Ramponi, P. Laporta, D. Polli, S. De Silvestri, and G. Cerullo,}
\newblock \bibinfo{journal}{J.  Opt. Soc.  Am. B}
\textbf{\bibinfo{volume}{20}}, 1559
(\bibinfo{year}{2003}).

%

\bibitem{gattass2008flm}
\bibinfo{author}{R. R. Gattass and E. Mazur,}
\newblock \bibinfo{journal}{Nature Photon.}
\textbf{\bibinfo{volume}{2}}, 219
 (\bibinfo{year}{2008}).


\bibitem{Spagnolo13} 
\bibinfo{author}{N. Spagnolo, C. Vitelli,	L. Aparo, P. Mataloni, F. Sciarrino, A. Crespi, R. Ramponi, and R. Osellame,}
\newblock {Nat. Commun.} {\bf 4}, 1606
(2013).

\bibitem{Crespi2016} 
\bibinfo{author}{A. Crespi, R. Osellame, R. Ramponi,  M. Bentivegna, F. Flamini,
N. Spagnolo, N. Viggianiello, L. Innocenti, P. Mataloni and F. Sciarrino}
\newblock {Nat. Commun.} {\bf 7}, 10469
(2016).




\bibitem{marshall_olkin}
\bibinfo{author}{A. W. Marshall and I. Olkin,}
\newblock \bibinfo{title}{\textit{Inequalities: Theory of Majorization and its Applications.}}
\newblock (\bibinfo{journal}{Acad. Press Inc., New York, 1979}).


\bibitem{vidal_cirac_02}
\bibinfo{author}{G. Vidal and J. I. Cirac,}
 \newblock \bibinfo{journal}{Phys. Rev. A}
 \textbf{\bibinfo{volume}{66}} (\bibinfo{year}{2002}).



\bibitem{grover_96}
\bibinfo{author}{L. K. Grover,}
\newblock \bibinfo{journal}{Phys. Rev. Lett.}
\textbf{\bibinfo{volume}{79}}, 325
(\bibinfo{year}{1997}).


\bibitem{shor_97}
\bibinfo{author}{P. W. Shor,}
\newblock \bibinfo{journal}{SIAM J. Sci. Stat. Comp.}
\textbf{\bibinfo{volume}{26}}, 1484
(\bibinfo{year}{1997}).


\bibitem{BV97}
\bibinfo{author}{E. Bernstein and  U. V. Vazirani,}
\newblock {SIAM J. Comp.} (\bibinfo{year}{1997}).


\bibitem{Farhi2000}
\bibinfo{author}{ E. Farhi, J. Goldstone, S. Gutmann, and M. Sipser,}
\newblock {arXiv:quant-ph/0001106v1}.


\bibitem{Childs2002}
\bibinfo{author}{A. M. Childs, R. Cleve, E. Deotto, E. Farhi, S. Gutmann, and D. A. Spielman,}
\newblock In \emph{\bibinfo{booktitle}{Proceedings of the 35th annual ACM symposium on Theory of computing}}
(\bibinfo{year}{STOC, 2003}).




\bibitem{Barak2007}
\bibinfo{author} {R. Barak and Y. Ben-Aryeh,}
\newblock \bibinfo{journal}{J.  Opt. Soc.  Am. B}
\textbf{\bibinfo{volume}{24}}, 231
(\bibinfo{year}{2007}).
  
\bibitem{Clements16}
\bibinfo{author} {W. R. Clements, P. C. Humphreys, B. J. Metcalf, W. S. Kolthammer, and I. A. Walmsley,} 
\newblock \bibinfo{journal}{arXiv:1603.08788v1}.
 


\bibitem{Broome2013}
\bibinfo{author}{M. A. Broome, A. Fedrizzi, S. Rahimi-Keshari, J. Dove, S. Aaronson, T. C. Ralph, and A. G. White,}
\newblock \bibinfo{journal}{Science}
\textbf{\bibinfo{volume}{339}}, 794
 (\bibinfo{year}{2013}).

\bibitem{Spring2013}
\bibinfo{author}{J. B. Spring, B. J. Metcalf, P. C. Humphreys, W. S. Kolthammer, X. Jin, M. Barbieri, A. Datta, N. Thomas-Peter, N. K. Langford, D. Kundys \textit{et al.}},
\newblock \bibinfo{journal}{Science} \textbf{\bibinfo{volume}{339}}, 798
(\bibinfo{year}{2013}).

\bibitem{Till2012}
\bibinfo{author}{M. Tillmann, B. Dakić,	R. Heilmann,	S. Nolte, A. Szameit, and P. Walther},
\newblock \bibinfo{journal}{Nature Photon.}
\textbf{\bibinfo{volume}{7}}, 540
(\bibinfo{year}{2013}).

\bibitem{Crespi2012}
\bibinfo{author}{A. Crespi, R. Osellame, R. Ramponi, D. J. Brod, E. F. Galv$\tilde{a}$o, N. Spagnolo, C. Vitelli, E. Maiorino, P. Mataloni, and  F. Sciarrino,}
\newblock \bibinfo{journal}{Nature Photon.}
\textbf{\bibinfo{volume}{7}}, 545
(\bibinfo{year}{2013}).

\bibitem{Spagnolo2013a}
\bibinfo{author}{N. Spagnolo, C. Vitelli,	M. Bentivegna, D. J. Brod, A. Crespi, F. Flamini, S. Giacomini, G. Milani, R. Ramponi, P. Mataloni, R. Osellame, E. F. Galv$\tilde{a}$o, and F. Sciarrino},
\newblock {Nature Photon.} \textbf{\bibinfo{volume}{8}}, 615
(\bibinfo{year}{2014}).

\bibitem{Carolan2013a}
\bibinfo{author}{J. Carolan, J. D. A. Meinecke, P. J. Shadbolt, N. J. Russell, N. Ismail, K. Wörhoff, T. Rudolph, M. G. Thompson, J. L. O'Brien, J. C. F. Matthews, and A. Laing},
\newblock {Nature Photon.} \textbf{\bibinfo{volume}{8}}, 621
(\bibinfo{year}{2014}).

\bibitem{Carolan2015}
\bibinfo{author}{J. Carolan, C. Harrold, C. Sparrow, E. Martin-Lopez, N. J. Russell, J. W. Silverstone, P. J. Shadbolt, N. Matsuda, M. Oguma, M. Itoh  \textit{et al.}},
\newblock {Science} {\bf 349}, 711
(\bibinfo{year}{2015}).

\bibitem{Bentivegna2015}
\bibinfo{author}{M. Bentivegna, N. Spagnolo, C. Vitelli,  F. Flamini, N. Viggianiello, L. Latmiral, P. Mataloni, D. J. Brod, E. F. Galv$\tilde{a}$o,  A. Crespi,  R. Ramponi,  R. Osellame and F. Sciarrino, }
\newblock \bibinfo{title}{Experimental scattershot boson sampling}
\newblock \emph{\bibinfo{journal}{Sci. Adv.}}
\textbf{\bibinfo{volume}{1}}, No. 3, (\bibinfo{year}{2015}).

\bibitem{AA10}
\bibinfo{author}{S. Aaronson and A. Arkhipov,}
\newblock In \emph{\bibinfo{booktitle}{Proceedings of the 43rd annual ACM symposium on Theory of computing}}
(\bibinfo{year}{ACM Press, 2011}).
  


  



\bibitem{AA13}
\bibinfo{author}{ S. Aaronson} and \bibinfo{author}{A. Arkhipov,}
\newblock \bibinfo{journal}{Quantum Inform. Compu.} \textbf{\bibinfo{volume}{14}}, 1383
(\bibinfo{year}{2014}).

\bibitem{Crespi2015} 
\bibinfo{author}{A. Crespi,}
\newblock {Phys. Rev. A} {\bf 91}, 013811 (2015).

\bibitem{Gogolin2015}
\bibinfo{author}{L. Aolita,	C. Gogolin, M. Kliesch and J. Eisert, }
\newblock \bibinfo{title}{Reliable quantum certification of photonic state preparations}
\newblock \emph{\bibinfo{journal}{Nat. Commun.}}
\textbf{\bibinfo{volume}{6}}, 8498, (\bibinfo{year}{2015}).

\bibitem{Friburgo2016}
\bibinfo{author}{M. Walschaers, J. Kuipers, J. D. Urbina, K. Mayer, M. C. Tichy, K. Richter and A. Buchleitner, }
\newblock \bibinfo{title}{Statistical benchmark for BosonSampling.}
\newblock \emph{\bibinfo{journal}{New J. Phys.}}
\textbf{\bibinfo{volume}{18}}, 032001, (\bibinfo{year}{2016}).

\bibitem{Spagnolo2016}
\bibinfo{author}{M. Bentivegna,  N. Spagnolo and  F. Sciarrino, }
\newblock \bibinfo{title}{Is my boson sampler working?}
\newblock \emph{\bibinfo{journal}{New J. Phys.}}
\textbf{\bibinfo{volume}{18(4)}}, 041001, (\bibinfo{year}{2016}).

\bibitem{Gogolin2013}
\bibinfo{author}{C. Gogolin, M. Kliesch, L. Aolita, and J. Eisert,}
\newblock \bibinfo{note}{arXiv:1306.3995v2.}



  

\bibitem{Scheel04} 
\bibinfo{author}{S. Scheel,}
\newblock {Acta Physica Slovaca} \textbf{58}, 675 (2008).



\bibitem{SI} 
See Supplemental Material for details on the quantum-to-classical transition and on the patterns of 2-photon distributions in Fourier interferometers.



\end{thebibliography}
\end{document}